# Mutation Sampling Technique for the Generation of Structural Test Data

M. Scholivé[1], V. Beroulle[1], C. Robach[1], M.L. Flottes[2], B.Rouzeyre[2]

[1] *LCIS-ESISAR, 50 rue B. de Laffemas, BP 54, 26902 Valence, France*

[2] *LIRMM, Université de Montpellier, 34090 Montpellier, France*

<Mathieu.scholive, Vincent.beroulle, Chantal.robach>@esisar.inpg.fr

<flottes, rouzeyre>@lirmm.fr

**Abstract**

*Our goal is to produce validation data that can be used as an efficient (pre) test set for structural stuck-at faults. In this paper, we detail an original test-oriented mutation sampling technique used for generating such data and we present a first evaluation on these validation data with regard to a structural test.*

## 1  Introduction

The presented approach addresses the data generation problem for both validation and physical tests. Data for validation are generated from high-level descriptions using a software testing technique. This paper focuses on the interest of re-using and optimizing these validation data for hardware testing purpose. As validation data are already generated when structural test generation begins, we propose to use them as a primary and "free" test set for structural faults. Obviously, to achieve very high fault coverage, this first test set only relying on the validation data will be completed with additional data obtained from a classical gate-level *Automatic Test Pattern Generation* (*ATPG*) process. Validation data reuse should decrease the gate-level test generation effort and final test application time.

This paper presents a software testing method for generating validation data. It is based on the well-known mutation testing principle and uses an original sampling technique for decreasing test generation time of these validation data without degrading validation results.

The sequel of this paper is organized as follows. Section 2 presents the classical mutation testing approach. In section 3, mutation operator efficiency is studied. Then, in section 4, we propose a new mutation sampling technique. The paper concludes with section 5.

## 2  Mutation Testing Overview

Originally proposed in [1] as a technique for unit software testing, the aim of mutation testing is to measure the efficiency of a test set to exercise the different functions of a program. This measure can also be used to generate test cases selecting only input data that are mutation adequate. It has been proved in [2] that the data generated by this approach meet most of design validation criteria such as statement coverage, branch coverage, ….

To generate validation data with mutation testing, we select vectors that can distinguish a program from a set of faulty versions of this program, the so-called *mutants*. These faults, i.e. "small" and syntactically correct modifications of the original instructions, are classified with the help of mutation operators. For VHDL descriptions, a set of ten operators has been defined in [3].

Through mutant simulation, this approach leads to a metric called the *Mutation Score* (*MS*). This metric measures the Test Set (*TS*) quality with respect to a program *P*. Before to define the *M*S, lets first define killed and equivalent mutants.

A *killed mutant* is a mutant that can be distinguished from the original program because it exists at least one data in TS that, when applied on inputs of the original program or the mutant, results in different outputs.

An equivalent mutant cannot be distinguished from the original program whatever the simulated input data.

The mutation score MS is computed as follow:

$$MS(TS,P) = \frac{K}{M - E}$$

Where *M* is the number of generated mutants,
*K* is the number of *killed mutants*,
E is the number of *equivalent mutants*.

## 3  Mutation Operator Efficiency

The mutation-based technique is a very time/memory consuming validation technique that must be optimized for complex circuits. A common strategy called mutation sampling consists in selecting a subset of mutants among the whole set of mutants generated from all the mutation operators. At the evidence, if we want to limit the generation effort performed at high level and re-use the validation sequence for a structural test, we must adjust our sampling strategy according to the fault coverage efficiency of the mutation operators. For this, we are going to select all the more mutants generated from one mutation operator than this operator is efficient with the regard to the stuck-at fault coverage.

Because validation data are considered as free data with regard to the detection of stuck-at faults, we compare the so generated test data with pseudo-random test sets generally used as initial test sets before to run an Automatic Test Pattern generation for hard to detect faults. We define metric, called the *Non Linear Fault Coverage Efficiency* (*NLFCE)* that allows to take into account the non-linear increasing difficulty to achieve high fault coverage level. This metric considers both the achieved fault coverage and the corresponding test length. First, stuck-at fault simulations performed with validation data on gate-level descriptions deliver corresponding fault coverages: the *Mutation Fault Coverage* (*MFC*). In the same way, fault simulations performed with pseudo-random test vectors allow to compute the *Random Fault Coverage* (*RFC*). *ΔFC%* is the relative fault coverage gain between mutation and random data for equal length sequences. *ΔL%* is the relative length gain between mutation and random data to achieve the same fault coverage. *NLFCE* is the product *ΔFC%.ΔL%*.

The experiments are performed on the *ITC'99 benchmarks* [4] and on the *ISCAS'85 benchmarks* [5]. Table 1 presents several results per mutation operator. Note that all mutation operators are not necessarily applied on every benchmark circuit. For instance, the *CR* (*Constant Replacement*) operator is only used if the high level description includes a constant declaration.

| Circuit | Operator | ΔFC% | ΔL% | NLFCE |
|---|---|---|---|---|
| **b01** | LOR | 0.66 | 10.84 | +7.16 |
|  | VR | 1.36 | 17.43 | +23.7 |
|  | CVR | 1.72 | 18.81 | +32.3 |
|  | CR | 2.32 | 37.60 | **+87.3** |
| **b03** | VR | 4.10 | 28.39 | +116 |
|  | CVR | 8.08 | 55.29 | +447 |
|  | CR | 9.57 | 49.89 | **+477** |
| **c432** | LOR | 4.14 | 32.35 | +134 |
|  | VR | 9.40 | 56.62 | +532 |
|  | CVR | 11.67 | 81.86 | **+955** |
| **c499** | LOR | 4.72 | 64.26 | +303 |
|  | VR | 6.18 | 73.10 | **+452** |
|  | CVR | 4.53 | 84.96 | +385 |

**Tab. 1: Operator Fault Coverage Efficiency**

These experimental results show that the *LOR* (*Logical Operator Replacement*) mutation operator is always the least efficient operator for stuck-at fault detection. Other operators can be ordered with regards to the efficiency (increasing order): *VR* (*Variable Replacement*), *CVR* (*Constant for Variable Replacement*) and *CR*. In other words, when the circuit descriptions include constant declarations, *CR* seems to be the most efficient operator. Obviously, this high level fault model is also well related to the stuck-at fault model.

## 4 Mutation Testing Strategy

The usual mutation sampling strategy [6] consists in sampling a low percentage of mutants, for instance 10%. Generally, this 10% are selected randomly. Our strategy consists in selecting the same final number of mutants (10% over the whole set of mutants) but this selection is not performed randomly. We select different percentages of mutants in the mutant subsets generated from different operators. The proportion of mutants selected from each operator is function of its stuck-at fault coverage efficiency.

Several experiments have been conducted on benchmark circuits for comparing the classical and the proposed sampling technique. Since the proposed strategy must preserve validation and structural test efficiencies, both *MS* (computed on all mutants) and *NLFCE* parameters are observed. These results are summarized in table 2. Obviously, the two strategies extract exactly the same percentage of mutants, which has been fixed to 10%.

|  | Test-oriented sampling 10% | | Random Sampling 10% | |
|---|---|---|---|---|
| Circuit | MS% | NLFCE | MS% | NLFCE |
| b01 | 85.98 | +340 | 83.71 | +278 |
| b03 | 64.16 | +1089 | 62.22 | +712 |
| c432 | 88.18 | +708 | 85.62 | +419 |
| c499 | 94.75 | +518 | 90.32 | +500 |

**Tab. 2: Our Testing Strategy Vs Mutant Sampling**

For instance, concerning the *c432* circuit, 77 mutants have been selected from the two strategies. Validation data are generated from this subset of mutants, and then applied to the entire population of mutants to provide the MS. With the classical random sampling technique, this MS equals to 85.62% and the *NLFCE* roughly equals to +400. With our sampling strategy, we increase the *MS* to 88.18%, and the *NLFCE* is roughly +700. Our strategy is thus more efficient for structural test comparing to the classical mutation sampling technique.

## 5 Conclusion

In this paper, we have proposed a strategy to reduce the simulation time preserving both validation and test efficiencies. This strategy consists in performing mutation sampling and is built thanks to the study of the efficiency of each mutation operator.

To demonstrate that validation data re-use leads to an efficient reduction in terms of ATPG effort, further experiments must be conducted on more complex designs.